\documentclass[ reprint,superscriptaddress, amsmath,amssymb,aps]{revtex4-1}
\usepackage{graphicx}
\pdfoutput=1
\usepackage{epstopdf}
\usepackage{color}
\usepackage{dcolumn}
\usepackage{bm}
\usepackage{hyperref}
\usepackage{upgreek} 
\usepackage{siunitx}
\usepackage{booktabs}
\usepackage[normalem]{ulem} 

\newcommand{\nPhot}{\ensuremath{n_{\upgamma}}}
\newcommand{\nElec}{\ensuremath{n_{\mathrm{e}}}}

\newcommand{\GeVmass}{\ensuremath{\mathrm{GeV}\,c^{-2}}}
\newcommand{\ra}[1]{\renewcommand{\arraystretch}{#1}} 

\begin{document}

\title{Results from a search for dark matter in the complete LUX exposure}

\author{D.S.~Akerib} 
\affiliation{Case Western Reserve University, Department of Physics, 10900 Euclid Ave, Cleveland, OH 44106, USA}
\affiliation{SLAC National Accelerator Laboratory, 2575 Sand Hill Road, Menlo Park, CA 94205, USA}
\affiliation{Kavli Institute for Particle Astrophysics and Cosmology, Stanford University, 452 Lomita Mall, Stanford, CA 94309, USA}

\author{S.~Alsum} 
\affiliation{University of Wisconsin-Madison, Department of Physics, 1150 University Ave., Madison, WI 53706, USA}

\author{H.M.~Ara\'{u}jo} 
\affiliation{Imperial College London, High Energy Physics, Blackett Laboratory, London SW7 2BZ, United Kingdom}

\author{X.~Bai} 
\affiliation{South Dakota School of Mines and Technology, 501 East St Joseph St., Rapid City, SD 57701, USA}

\author{A.J.~Bailey} 
\affiliation{Imperial College London, High Energy Physics, Blackett Laboratory, London SW7 2BZ, United Kingdom}

\author{J.~Balajthy} 
\affiliation{University of Maryland, Department of Physics, College Park, MD 20742, USA}

\author{P.~Beltrame} 
\affiliation{SUPA, School of Physics and Astronomy, University of Edinburgh, Edinburgh EH9 3FD, United Kingdom}

\author{E.P.~Bernard} 
\affiliation{University of California Berkeley, Department of Physics, Berkeley, CA 94720, USA}
\affiliation{Yale University, Department of Physics, 217 Prospect St., New Haven, CT 06511, USA}

\author{A.~Bernstein} 
\affiliation{Lawrence Livermore National Laboratory, 7000 East Ave., Livermore, CA 94551, USA}

\author{T.P.~Biesiadzinski} 
\affiliation{Case Western Reserve University, Department of Physics, 10900 Euclid Ave, Cleveland, OH 44106, USA}
\affiliation{SLAC National Accelerator Laboratory, 2575 Sand Hill Road, Menlo Park, CA 94205, USA}
\affiliation{Kavli Institute for Particle Astrophysics and Cosmology, Stanford University, 452 Lomita Mall, Stanford, CA 94309, USA}

% ...

\collaboration{LUX Collaboration}
\date{\today}
\begin{abstract}
\vspace*{-4mm}
We report constraints on spin-independent weakly interacting massive particle (WIMP)-nucleon scattering using a $3.35\times10^4$\,kg\,day exposure of the Large Underground Xenon (LUX) experiment. A dual-phase xenon time projection chamber with 250\,kg of active mass is operated at the Sanford Underground Research Facility under Lead, South Dakota (USA). With roughly fourfold improvement in sensitivity for high WIMP masses relative to our previous results, this search yields no evidence of WIMP nuclear recoils.  At a WIMP mass of 50\,GeV\,$c^{-2}$, WIMP-nucleon spin-independent cross sections above 2.2$\times10^{-46}$\,cm$^2$ are excluded at the 90\% confidence level.  When combined with the previously reported LUX exposure, this exclusion strengthens to 1.1$\times10^{-46}$\,cm$^2$ at 50\,GeV\,$c^{-2}$.
\end{abstract}
\maketitle
The Large Underground Xenon (LUX) experiment searches for direct evidence of weakly interacting massive particles (WIMPs), a favored dark matter candidate. The LUX search is performed with a dual-phase (liquid-gas) xenon time projection chamber (TPC) containing 250\,kg of ultrapure liquid xenon (LXe) in the active detector volume\,\cite{Akerib:2012ak}. Energy deposited by particle interactions in the LXe induces two measurable signal channels: prompt VUV photons from scintillation (S1), and free electrons from ionization.  The S1 photons are emitted from the interaction site and detected by top and bottom arrays of photomultiplier tubes (PMTs). Electrons liberated by the interaction drift to the surface of the liquid via an applied electric field. They are extracted into the gas and accelerated by a larger electric field, producing secondary electroluminescence photons collected in both arrays with localization in the top PMTs (S2). The PMT signals from both light pulses, S1 and S2, allow for the reconstruction of interaction vertices in three dimensions.

The ratio of the S1 and S2 signals is used to discriminate between electronic recoils (ER) and nuclear recoils (NR). WIMP interactions in the detector would primarily appear as nuclear recoils of energy~$\lesssim$~100\,keV\,\cite{Lewin199687}. In order to reduce backgrounds from external sources, the detector is immersed in a 7.6\,m diameter and 6.1\,m high ultrapure water tank, which itself is located underground at the Sanford Underground Research Facility (SURF) in Lead, SD, USA. The $\sim$1.5\,km of rock overburden (4300 m.w.e.) provides a reduction in the rate of cosmic muons of $\mathcal{O}(10^{-7})$. ER background populations arise from $^{40}$K and the $^{238}$U/$^{232}$Th decay chains present as contaminants in materials other than LXe, as well as from trace amounts of $^{222}$Rn and $^{85}$Kr in the LXe itself. The $^{85}$Kr is largely removed from the xenon prior to filling by chromatographic separation in activated charcoal\,\cite{Akerib:krremoval}. Additional information on the experimental setup\,\cite{Akerib:2011:daq, Akerib:2015:trigger, Akerib:2012ys} and backgrounds\,\cite{Akerib:2014:bg} has been previously published.

The first LUX WIMP search (WS2013) collected 95 live-days of data from April to August, 2013 \,\cite{Akerib:2013:run3, Akerib:2015:run3, Akerib:2016:SD}. Extensive periods of calibration under the same WS2013 running conditions followed, including NR calibrations using neutrons from a deuterium-deuterium (DD) beam\,\cite{Akerib:2015:dd, Verbus:2016sgw}, and low-energy ER calibrations using $^3$H beta decay\,\cite{Akerib:2015:tritium}. This novel calibration program has markedly extended the understanding of the LXe detection medium for low-energy interactions; the sensitivity of the WIMP searches has consequently improved, particularly for low-mass WIMPs.

In preparation for the WIMP-search exposure reported here (WS2014--16), the anode, gate, and cathode grid electrodes underwent a campaign of ``conditioning'' in cold Xe gas, during which each electrode's applied voltage was elevated just above the onset of sustained discharge and maintained for a multiday period, akin to the burn-in period often employed in room-temperature proportional counter commissioning\,\cite{conditioning1, conditioning2, conditioning3, conditioning4, conditioning5}. The goal of this campaign was to improve the voltages at which the electrodes could be biased. As a result, the measured electron extraction efficiency (i.e.~the fraction of electrons which promptly cross the liquid--gas interface) increased from (49$\pm$3)\% in WS2013 to (73$\pm$4)\% in WS2014--16. Following the conditioning campaign and extensive calibrations at the new operating voltages, WS2014--16 ran from September 11, 2014 until May 2, 2016, during which time the detector conditions were kept uniform. The long-term behavior of the LXe temperature and pressure varied by less than 0.5\,K and $10^{-2}$\,bar. The electron lifetime in the LXe was typically stable for long durations and above 1\,ms, significantly longer than the maximum electron drift time of $\sim$400\,$\mu$s. Periods of low ($<$500\,$\mu$s) electron lifetime are excluded from this analysis (including an extended period from March 24 to June 2, 2015), as were periods in which detector-stability parameters (e.g.~pressure, temperature, liquid level, recirculation flow rate) deviated by more than a few percent over short time scales. The WS2014--16 exposure consists of 332.0 live days.

Though the grid conditioning campaign achieved the goal of an increased electron extraction efficiency, it was observed during calibrations that electron drift trajectories were significantly altered from the near-vertical paths seen in WS2013. In WS2013, due to field cage geometry alone (similar to \,\cite{xenon100:fieldthesis}), electrons emitted near the periphery of the cathode grid, at a starting radius of $\sim$24\,cm, were directed slightly radially inwards during their upward drift, exiting the liquid surface at an S2 radius ($\mathrm{r_{S2}}$) of $\sim$20\,cm. In WS2014--16, a stronger radial effect is seen. Electrons of the same cathode-edge starting radius ($\sim$24\,cm) exit the liquid surface at $\sim$10\,cm; the strength of this effect varies with both azimuth and date. These observations are consistent with a nonuniform and time-varying negative charge density in the polytetrafluoroethylene (PTFE) panels which define the radial boundary of the active volume. This PTFE charge is understood as resulting from exposure to coronal discharge during the grid conditioning. The VUV photons produced in this process can liberate PTFE electron-hole pairs. As the holes in PTFE have a significantly higher mobility than the electrons\,\cite{ptfe_bandgap,ptfe_trapdepths}, the applied electric field preferentially removes holes, resulting in a buildup of net negative charge over long time scales.  The observed charge densities and transport time scales are consistent with values in the literature\,\cite{ptfe_trapdensity,ptfe_chargetimecales}.

A time-dependent mapping between true recoil position and the ``observed S2 coordinates'' of $\mathrm{\{x_{S2}}$, $\mathrm{y_{S2}}$, and drift time $\mathrm{z_{S2}}\}$ is required for interpretation of the data, necessitating the construction of an electric field model. The \textsc{comsol Multiphysics} package\,\cite{comsolRef} is used to build a 3-D electrostatic model of the LUX detector, including a heterogeneous and date-specific charge density in the PTFE panels. This charge density is fit to data from regular ($\sim$weekly) $^{83\mathrm{m}}$Kr\,\cite{Kastens:2009,Kastens:JINST,Manalaysay:2009yq} calibrations, each producing $\sim$10$^6$ events of uniform true recoil position within the active volume. The heterogeneous PTFE charge density is modeled by dividing the PTFE surface into a grid of 42 sections (seven sections in height, six in azimuth),  each section having a variable uniform charge density. These 42 individual electrostatic charge densities are varied through a Metropolis-Hastings Markov Chain Monte Carlo algorithm fitting procedure\,\cite{chib:95,Saha:02}, minimizing the difference in $\mathrm{\{x_{S2}}$, $\mathrm{y_{S2}}$, $\mathrm{z_{S2}}\}$ distribution boundaries between simulation and data. Field and charge maps are updated on a monthly basis, although the variation time scale is observed to be longer. The average PTFE charge density is observed to increase in magnitude over the course of the exposure, starting at $-3.6$\,$\mu$C/m$^2$ and asymptotically approaching $-5.5$\,$\mu$C/m$^2$.
In the WIMP-search analysis, comparisons of data to models of signal and background are most naturally performed in the observed S2 coordinates of $\mathrm{\{ x_{S2} , y_{S2} , z_{S2} \} }$. Data are kept in these observed S2 coordinates, while the true recoil positions of simulated data are mapped into this space using field models mentioned above. Comparisons between the observed and modeled spatial distributions (see Appendix\,\ref{sec:appendix_section}), discussed later, show excellent agreement.

A generic feature of dual-phase TPCs is that measured S1 and S2 signals from a monoenergetic source will vary according to the vertex position of the interaction. For S1, this is due to spatially varying geometrical conditions that affect the efficiency for detecting S1 photons. In LUX, this detection efficiency is larger for photons emitted close to the cathode, and smaller for photons emitted close to the liquid surface (a variation of around 30\%). For S2, a similar variation results from the loss of electrons to electronegative impurities in the LXe (a date-dependent variation of around 20\%--50\%). The variations in S1 and S2 due to these geometrical effects are independent of the incident particle type and deposited energy.  In WS2013, a position-dependent correction map for these effects was derived in a straightforward manner, by measuring the spatial variation in S1 and S2 from a monoenergetic $^{83\mathrm{m}}$Kr calibration source.

In WS2014--16, this picture is complicated because the spatially varying electric field magnitude influences the recombination of electron-ion pairs, changing the yields of photons and electrons emitted at an interaction vertex before the geometrical effects come into play. As the electric field magnitude is increased, fewer photons and more electrons escape the interaction\,\cite{Aprilenc:2006}. 
For the 50 to 600\,V/cm field variation over the fiducial region relevant to this analysis, the average light yield for a 5\,keV ER event falls by 15\%, while average charge yield rises by the same amount. The scale of variation is less pronounced for lower-energy ER events\,\cite{Mock:2014,Lin:2015jta}. 
For a 5\,keV NR event, the field-induced changes in light and charge yield are smaller, at the level of 5\%\,\cite{Lenardo:2014}. The observed total spatial variation in S1 and S2 from a monoenergetic calibration source is therefore a combination both of field effects and geometrical effects. The geometrical effects are independent of particle type and energy deposition, but the field effects depend strongly on these factors. Therefore, a position-dependent correction map can only be universally applied to all observed signals if it corrects for geometrical effects only. 

Several techniques are employed to separate the geometrical effects from the field effects, enabling the desired correction of observed signals for geometrical effects alone. The field effects remain in the observed science data, and are similarly included in the background and signal models for interpretation.

Two calibration tools enable the construction of geometry-only correction maps. The first is $^{83\mathrm{m}}$Kr, which decays in two steps: 32.1\,keV and 9.4\,keV. These steps are separated by a decay constant of 154\,ns, thereby producing two S1 pulses. While the variation in size of these S1 pulses depends on several factors, the variation in the ratio of the two depends only on the applied field\,\cite{Manalaysay:2009yq}. Second, the field effect for low-energy electronic recoils is extremely weak\,\cite{Szydagis:2011}. Observed variations in the position of the $^3$H spectral maximum (2.5\,keV) are therefore almost entirely due to geometrical effects alone. Leveraging the $^{83\mathrm{m}}$Kr S1 ratio that depends on field alone, and low-energy $^3$H response that depends nearly on geometry alone, geometry-only corrections are constructed.

The italicized quantities \textit{S1} and \textit{S2} indicate signal amplitudes that have been corrected for geometrical effects; \textit{S1} is normalized to the center of the active xenon, while \textit{S2} is normalized to the top of the active xenon. Using these quantities, gain factors $g_1$ and $g_2$ are defined through the expectation values $\langle\textit{S1}\,\rangle=g_1\nPhot$ and $\langle\textit{S2}\,\rangle=g_2\nElec$, given \nPhot\ initial photons and \nElec\ initial electrons leaving the interaction site. The $g_1$ and $g_2$ values in WS2014--16 are found using a set of monoenergetic electronic-recoil sources as in\,\cite{Akerib:2015:run3}, and are observed to vary slightly over the course of the exposure, independent of the field variation. The $g_2$ value varies within the range of 18.92\,$\pm$\,0.82 to 19.72\,$\pm$\,2.39~phd per liquid electron; $g_1$ gradually falls from 0.100\,$\pm$\,0.002 to 0.097\,$\pm$\,0.001~phd per photon. Here, ``phd'' indicates ``photons detected,'' differing from the more commonly used unit of photoelectrons (phe) through a small factor representing the probability of a single photon to produce multiple phe in a PMT cathode\,\cite{Faham:2015kqa}. Using $\hat{n}_{\gamma}\equiv\textit{S1}/g_1$ and $\hat{n}_{\mathrm{e}}\equiv\textit{S2}/g_2$, the ER combined energy scale (CES) is constructed as $E_{\textsc{ces}} \equiv (\hat{n}_{\gamma} + \hat{n}_{\mathrm{e}})\times 13.7$\,eV\,\cite{Shutt:2006ed}. This observable is independent of electric field because of the experimentally observed anticorrelation of $\nPhot$ and $\nElec$\,\cite{Conti:2003av,Aprile:2007qd,attila:thesis}.  
Spatial variations in the $E_{\textsc{ces}}$ peak position of a monoenergetic source are therefore due to geometrical effects only, and are used as a cross-check to verify the accuracy of the geometry-only corrections. For all dates during the WS2014--16 run, the $E_{\textsc{ces}}$ peak position of $^{83\mathrm{m}}$Kr (41.5\,keV) varies by less than 1\% within the fiducial volume. A cross-check using the $E_{\textsc{ces}}$ peak of $^{131\mathrm{m}}$Xe (164\,keV) gives a spatial variation of 1.8\%.

\begin{figure}[ht!]
\begin{center}
\includegraphics[width=0.48\textwidth,clip]{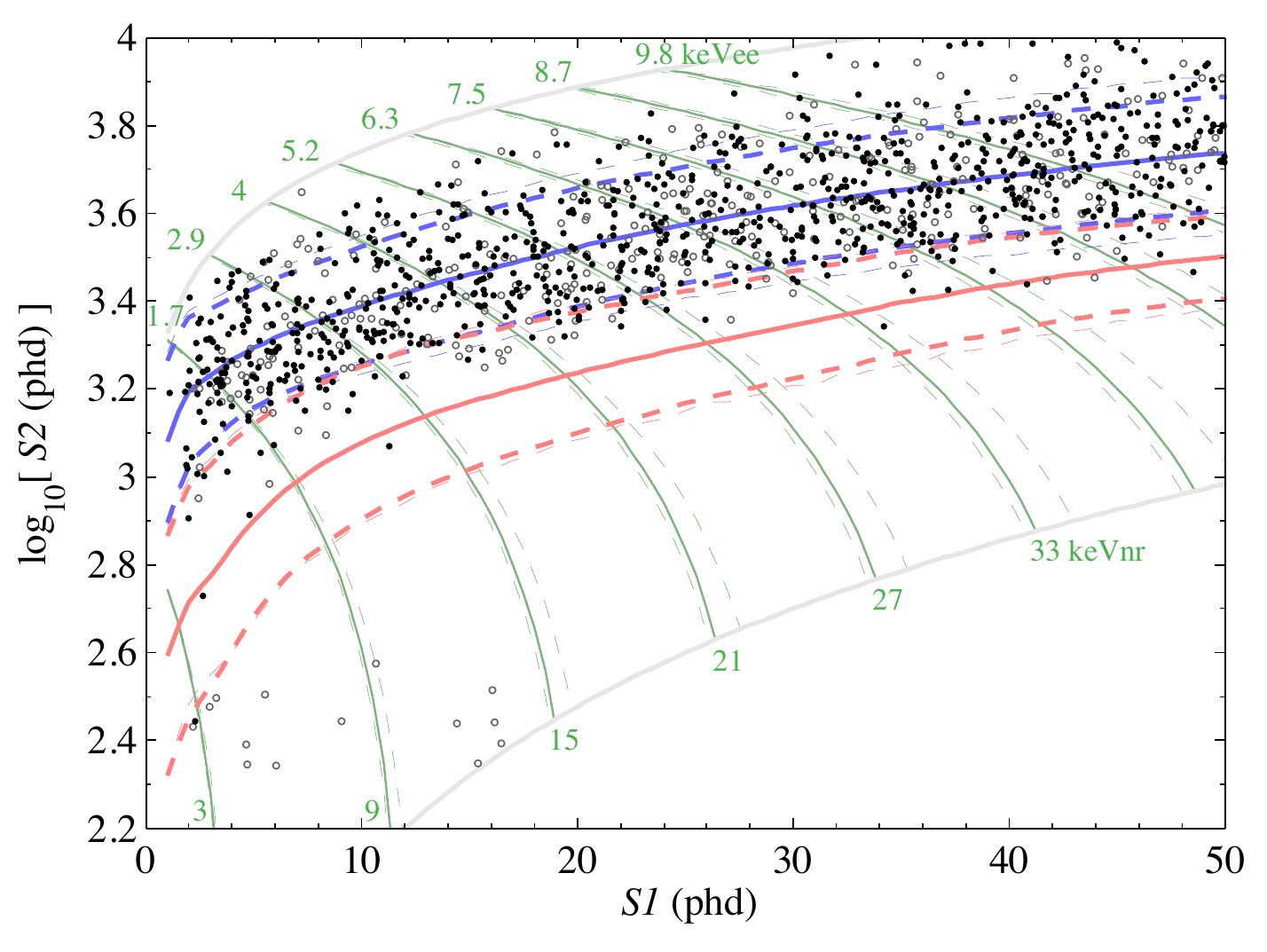}
\caption{WS2014--16 data passing all selection criteria. Fiducial events within 1 cm of the radial fiducial volume boundary are indicated as unfilled circles to convey their low WIMP-signal probability relative to background models (in particular the $^{206}$Pb wall background). Exposure-weighted average ER and NR bands are indicated in blue and red, respectively (mean, 10$\%$, and 90$\%$ contours indicated).  Of the 16 models used, the scale of model variation is indicated by showing the extrema boundaries (the upper edge of the highest-S2 model and the lower edge of the lowest-S2 model) as fainter dashed lines for both ER and NR. Gray curves indicate a data selection boundary applied before application of the profile likelihood ratio method. Green curves indicate mean (exposure-weighted) energy contours in the ER interpretation (top labels) and NR interpretation (lower labels), with extrema models dashed.}
\label{fig:background_data}
\end{center} 
\end{figure}

The electric field dependencies of \emph{S1} and \emph{S2} yields are included in the analysis by dividing the WS2014--16 exposure into ``exposure segments'', each having its own ER and NR detector-response model. There are 16 such segments, constructed by dividing the exposure into four bins of drift time (related to event depth) and four bins of date. Within each exposure segment, the field magnitude is considered to be constant and uniform. Boundaries in date are September 11, 2014; January 1, 2015; April 1, 2015; October 1, 2015; May 2, 2016. Boundaries in drift time are 40, 105, 170, 235, 300\,$\mu$s. Periodic $^3$H calibrations provide each of the 16 exposure segments with a unique calibration set from which to construct a unique individual response model. These 16 response models take the form of parameter variations of the Noble Element Simulation Technique (NEST) model\,\cite{Lenardo:2014}, which captures both the LXe microphysics of signal production and the detector physics of signal collection. Fits are performed by comparing the measured ER band (median and 10--90 percentile width in the $\mathrm{\{ \emph{S1}, \log_{10}(\emph{S2})\}}$ plane as in Fig.~\ref{fig:background_data}) with that predicted by the response model, in the range of 0--50\,phd, which roughly corresponds to an energy range of 0--10\,keVee. Specific to each exposure segment, two model parameters are varied during these fits: the electric field magnitude, and the recombination fluctuation parameter $F_r$ (see\,\cite{Szydagis:2011,Szydagis:2013,Mock:2014,Lenardo:2014}). Parameters that describe the detector as a whole (e.g.,~$g_1$, electron extraction efficiency, and S2 gas gain), are allowed to vary while constrained to be equal for all exposure segments within a given date bin. In each exposure segment, the measured ER band median differs from the model band median by less than 1\% for all \emph{S1}. The 16 electric field magnitudes found through these fits are consistent with the values earlier obtained from the electrostatic field models. This last point deserves emphasis, because the two techniques for estimating electric field magnitude are completely independent: the electrostatic field model is based on the observed electron drift paths alone, while the NEST fits are based on the \textit{S1} and \textit{S2} amplitudes alone. 

Neutron calibrations with the DD source were performed in each date bin.  For each individual exposure segment, the best-fit parameters from the corresponding ER calibration are applied to the NEST NR model.  The resulting NR models show excellent agreement with calibrations, such that the NR band medians of corresponding models and calibrations differ by less than 2.6\% for all \textit{S1}.  As in\,\cite{Akerib:2015:run3}, the overall energy scale in the response models is fixed by fitting the NEST NR model to a separate \textit{in situ} energy calibration using tagged neutron multiple scatters\,\cite{Akerib:2015:dd, Verbus:2016sgw}.  As before, we conservatively assume NR light yield to be zero below 1.1\,keV, the lowest energy at which NR light yield was measured in\,\cite{Akerib:2015:dd}. The 16 ER and 16 NR models are then used within a profile likelihood ratio (PLR) method\,\cite{Cowan:2010js} to search for evidence of dark-matter scattering events. It can be seen from the light-dashed curves in Fig.~\ref{fig:background_data}, representing extrema of the 16 ER and NR models, that the scale of model variation is small and diminishes towards the energy threshold.

Events consisting of a single scatter within the active LXe are selected according to several criteria: a single S2 preceded by a single S1, an S1 threshold of 2 PMT coincidence, and an upper threshold for the summed pulse area outside S1 and S2 within the trigger window. This last selection removes triggers during high single-extracted-electron activity following large-S2 events  \cite{Akerib:2015:run3,Chapman:thesis}, and results in 99.0$\%$ efficiency when applied to $^{3}$H calibration data for WS2014--16. The S2 threshold is set to 200\,phd (raw uncorrected pulse area) to avoid events for which the $\mathrm{\{x_{S2},y_{S2}\}}$ position uncertainty is high. Events for which \emph{S2}\,$>10^4$\,phd, \emph{S1}\,$>50$\,phd,  $\log_{10}(S\emph{2})<\,$median$_{NR}-5\sigma_{NR}$ or $\log_{10}(S\emph{2})>$\,median$_{ER}+3\sigma_{ER}$ (boundaries evident in Fig.~\ref{fig:background_data}) are considered far from the region of interest and are ignored.

A fiducial volume in drift time is defined as 40--300\,$\mu$s (date-independent). Each of the four date bins has a uniquely defined radial fiducial selection boundary, 3.0\,cm radially inward from the measured PTFE surface position for that date bin in observed S2 coordinates, $\mathrm{\{ x_{S2} , y_{S2} , z_{S2} \} }$. The wall position, a function of $\mathrm{\{ \phi_{S2} , z_{S2} \} }$, is measured with $^{210}$Pb subchain events that originate on the PTFE surface. The fiducial mass is determined by scaling the 250\,kg of active LXe by the acceptance fraction of $^{83\mathrm{m}}$Kr events through the fiducial-selection criteria. The time-averaged fiducial masses for the date bins are 105.4, 107.2, 99.2, and 98.4\,kg, in chronological order. A 3\% systematic uncertainty across all dates is estimated through comparison with acceptance fractions of $^3$H calibration data, of similarly uniform distribution in true recoil position. 

\begin{figure}
\begin{center}
\includegraphics[width=0.48\textwidth,clip]{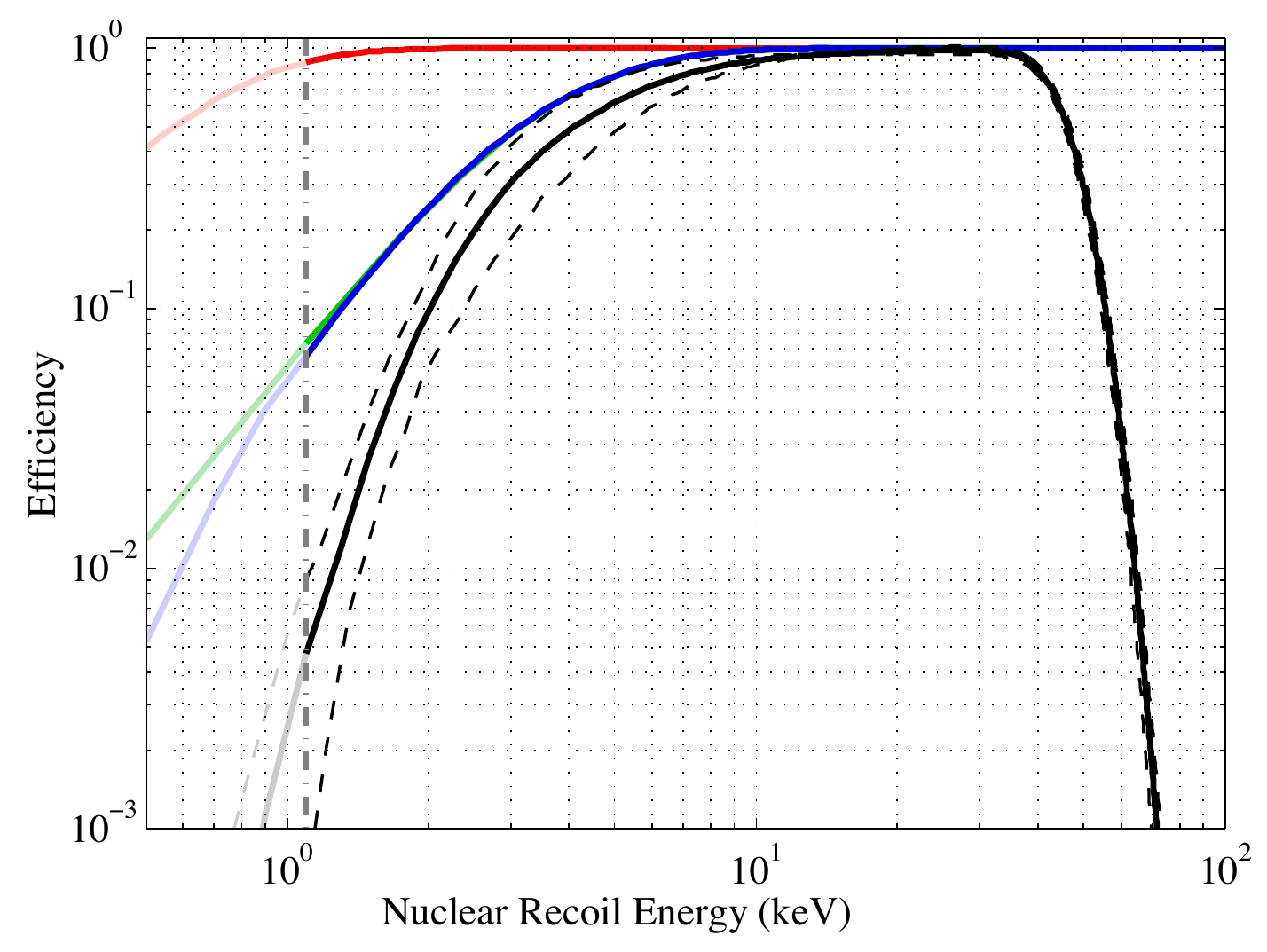}
\caption{Efficiencies for NR event detection, estimated using simulation with parameters tuned to calibration data. In descending order of efficiency---red: detection of an S2 (and classification as such by analysis); green: detection of an S1 ($\geq$2 PMTs detecting photons); blue: detection of both an S1 and an S2; black: detection passing analysis selection criteria. Solid curves indicate exposure-weighted means of the 16 calibrated models. The scale of model variation is illustrated by including the efficiencies of the date and $z$ bins with highest and lowest total efficiency (black dashed curves). Below 1.1\,keV nuclear recoil energy, the lowest energy for which light yield was measured in\,\cite{Akerib:2015:dd}, efficiency is conservatively assumed to be zero.}
\label{fig:efficiency}
\end{center}
\end{figure}

We apply additional pulse-quality cuts to eliminate pathological pulses which would otherwise be incorrectly classified as single-scatter interactions. The first of these populations is a class of energy depositions in the gaseous xenon (``gas events''), in which the entire gas event is classified as an S2 pulse. A cut targeting these pulses is formed by requiring $\sigma_{\mathrm{S}2}>0.4$\,$\mu$s, where $\sigma_{\mathrm{S}2}$ is the width resulting from a Gaussian fit to the pulse waveform. This cut has an acceptance of $>$90\% at the S2 threshold of 200\,phd, rising to $>$99\% for S$2>800$\,phd. The second pathological population is events in which two S2 pulses occur close together and are classified as a single S2 pulse (``merged multiple scatters''). Merged multiple scatters are rejected with cuts on the $\mathrm{\{x_{S2},y_{S2}\}}$ position reconstruction goodness-of-fit (this cut flagging multiple scatters separated in $x,y$) and on the ratio of $\sigma_{\textrm{S}2}$ to the time between the cumulative $1\%$ and $50\%$ area fraction (this cut flagging vertices separated in $z$). The combined efficiency for these cuts, calculated by applying these cuts to a population of known single-scatter $^{3}$H S2 waveforms, is $>$70\% at the 200\,phd S2 threshold, rising to $>$95\% for detected S$2>1000$\,phd. A summary of all efficiencies is shown as a function of NR energy in Fig.~\ref{fig:efficiency}.

For implementation in the PLR, a background model consisting of three classes of events is constructed: events of typical LXe charge and light yield, events affected by proximity to the PTFE surface, and accidental coincidences of isolated S1 and S2 pulses.

A background model representing recoils of typical charge and light yield is constructed much as in\,\cite{Akerib:2015:run3}. A counting of detector materials\,\cite{Akerib:2014:bg} informs a \textsc{Geant4}-based \textsc{LUXSim}\,\cite{Akerib:2012:luxsim} simulation. Two types of ER background populations are simulated: Compton scattering of $\upgamma$ rays (originating in trace radioactivity in detector components), and $\upbeta$ decays (originating in the bulk LXe from trace amounts of $^{85}$Kr and $^{222}$Rn daughters). Simulated true recoil positions are converted to S2 observed coordinates $\mathrm{\{ x_{S2} , y_{S2} , z_{S2} \} }$ using electric field maps specific to each date bin. Distributions in $\mathrm{\{ \emph{S1}, \log_{10}(\emph{S2})\}}$ result from the NEST model specific to the simulated exposure segment. The contributions of these ER backgrounds are additionally constrained by the WIMP-search data, selecting a region of the ER band [$\log_{10}(S\emph{2})>\, $median$_{ER}$] that avoids overlap with the NR signal region. There are two NR background event populations: neutrons (from detector components and cosmic muons), and coherent elastic nuclear scatters of $^8$B solar neutrinos. Single-scatter neutron interaction rates have been estimated through radioactivity screening data, simulations, and tests for multiple scatter neutron events. Simulations show that the multiple scatter event rates are significantly higher than the single scatter rates, and so the former can be used to establish upper limits on the latter event rates. These analyses show that single scatter neutron events can be left out of the background model due to their negligible event contribution in the WS2014--16 exposure. The $^8$B solar neutrino background is included as a low-rate NR background contribution in the PLR model.

Events from radon progeny on the PTFE surface can exhibit suppressed charge yield, due to charge loss to the PTFE (some radon progeny exhibit further charge suppression due to nuclear recoil type, as in $^{210}$Po decay, emitting $^{206}$Pb nuclei). The true recoil positions of these events are $\ll$1~mm from the wall surface, and as a result inward leakage from the wall surface in the $\mathrm{\{ r_{S2} , \phi_{S2} , z_{S2}\}}$ observation space is determined by S2 position reconstruction uncertainty alone. This uncertainty scales as $S\emph{2}^{-1/2}$. A small fraction of these events can leak into the fiducial volume near the S2 threshold. This population at high radius and low log$_{10}S\emph{2}$ can be seen in Fig.~\ref{fig:background_data}. An empirical model is constructed similar to\,\cite{Akerib:2015:run3}, using two samples of the WIMP-search data outside the region of interest. The PDF in $\mathrm{\{} S\emph{1}  , \log_{10}(S\emph{2}) ,\mathrm{ \phi_{S2} , z_{S2}\}}$ is inferred from a high-radius sample (greater than 1\,cm beyond the fiducial boundary). A high-S1 sample ($\emph{S1} > 55$\,phd) of events below the NR median is used to characterize the radial distribution of these events as a function of \emph{S2}.

Isolated S1 pulses appear in the event record, as do isolated S2 pulses. Though these pulses are rare, they may accidentally occur close enough in time (and in the correct order) to resemble a single-scatter energy deposition in the LXe. The $\mathrm{\{} S\emph{1},\,\log_{10}(S\emph{2})\mathrm{\}}$ distribution of these accidental coincidences, $f_{\mathrm{acc}}$, is taken to be separable, that is, $f_{\mathrm{acc}}(S\emph{1},\,\log_{10}\! S\emph{2}) = f_1(S\emph{1})\times f_2(\log_{10}S\emph{2})$.  The individual differential rates of isolated S1 pulses ($f_1$) and isolated S2 pulses ($f_2$) are measured from WIMP-search data. Because of their uncorrelated nature, these events are modeled as uniform in $\mathrm{\{ x_{S2} , y_{S2} , z_{S2} \} }$.

A protocol for blinding the data to potential NR WIMP signatures, to reduce analysis bias, began on December 8th, 2014 and was carried through the end of the exposure. Artificial WIMP-like events (``salt'') were manufactured from sequestered $^3$H calibration data and introduced into the data at an early stage in the data pipeline, uniform in time and position within the fiducial volume. Individual S1 and S2 waveforms from this data set were paired to form events consistent with a nuclear recoil S2 vs S1 distribution. Some S2-only salt events were added as well. The nuclear recoil energy distribution of these events had both an exponential (WIMP-like) and flat component. The four parameters describing these distributions (the exponential slope, the flat population's end point, the total rate, and the relative ratio of exponential vs. flat rates) were chosen at random within loose constraints and were unknown to the data analyzers. The salt event trigger times were sequestered by an individual outside the LUX collaboration until formally requested for unblinding, after defining the data selection criteria, efficiencies, and PLR models.

Following the removal of salt events, two populations of pathological S1+S2 accidental coincidence events were identified in which the S1 pulse topologies were anomalous. In the first of these rare topologies, $\sim$80\% of the collected S1 light is confined to a single PMT, located in the edge of the top PMT array. This light distribution is inconsistent with S1 light produced in the liquid, but is consistent with light produced outside the field cage and leaking into the TPC. A loose cut on the maximum single PMT waveform area as a fraction of the total S1 waveform area is tuned on ER and NR calibrations to have $>$99$\%$ flat signal acceptance. The second population of anomalous events also features a highly clustered S1 response in the top array, as well as a longer S1 pulse shape than typical of liquid interactions; these pulses are consistent with scintillation from energy deposited in the gaseous xenon. A loose cut on the fraction of detected S1 light occurring in the first 120~ns of the pulse is similarly tuned on ER and NR calibration data to have $>$99$\%$ signal acceptance across all energies. These two cuts, developed and applied after unblinding, feature very high signal acceptance, are tuned solely on calibration data, and only eliminate events that clearly do not arise from interactions in the liquid.

The result presented here includes the application of these two postunblinding cuts, and additionally includes 31.82 live days of nonblinded data, collected at the beginning of the WS2014--16 exposure before the start of the blinding protocol.

\begin{table}[b]
\ra{1.2}
\caption{\label{tab:fit_parameters}Model parameters in the best fit to WS2014--16 data for an example 50\,\GeVmass\ WIMP mass. Constraints are Gaussian with means and standard deviations indicated. Fitted event counts are after cuts and analysis thresholds. 
}
\begin{tabular}{@{}l@{}r@{}lr@{}l@{}}
Parameter & Con & straint & Fit & \:Value\\ 
\hline
Lindhard $k$\,\cite{Akerib:2015:dd}& 0.174 & $\,\pm\, 0.006$ &  & - \\
Low-$z$-origin $\upgamma$ counts & 94 & $\,\pm\, 19$& $99$&$\,\pm\, 14$ \\
Other $\upgamma$ counts & 511 &$\,\pm\, 77$ & 590 & $\,\pm\, 34$ \\
$\upbeta$ counts & 468 & $ \,\pm\, 140$ & 499 & $\,\pm\, 39$ \\
$^{8}\textrm{B}$ counts & 0.16 & $\,\pm\, 0.03$ & $0.16 $ & $\,\pm\, 0.03$ \\
PTFE surface counts & $14 $ & $ \,\pm\, 5$& 12 & $\,\pm\, 3$ \\
Random coincidence counts & 1.3 & $\,\pm\, 0.4$ & $1.6$ & $\,\pm\, 0.3$ \\
\hline
\end{tabular}
\end{table}

\begin{figure}
\begin{center}
\includegraphics[width=0.48\textwidth,clip]{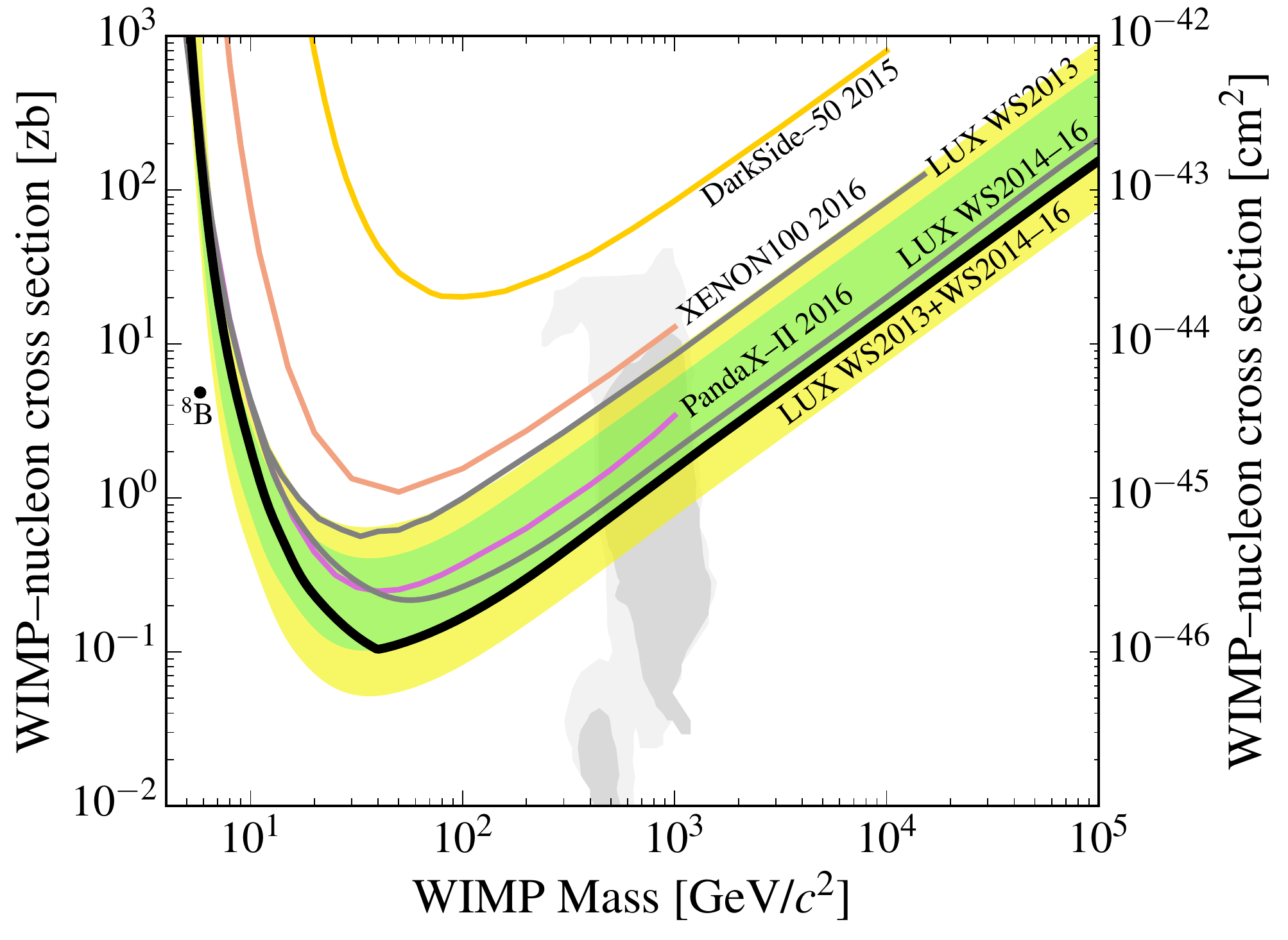}
\caption{Upper limits on the spin-independent elastic WIMP-nucleon cross section at 90\% C.L. The solid gray curves show the exclusion curves from LUX WS2013 (95 live days)\,\cite{Akerib:2015:run3} and LUX WS2014--16 (332 live days, this work).  These two data sets are combined to give the full LUX exclusion curve in solid black (``LUX WS2013+WS2014--16''). The 1-- and 2--$\sigma$ ranges of background-only trials for this combined result are shown in green and yellow, respectively; the combined LUX WS2013+WS2014--16 limit curve is power constrained at the --1$\sigma$ level. Also shown are limits from XENON100\,\cite{Aprile:2016swn} (red), DarkSide-50\,\cite{Darkside:2015} (orange), and PandaX-II\,\cite{Pandax:2016} (purple). The expected spectrum of coherent neutrino-nucleus scattering by $^{8}$B solar neutrinos can be fit by a WIMP model as in\,\cite{Billard:2013}, plotted here as a black dot. Parameters favored by SUSY CMSSM\,\cite{Bagnaschi:2015} before this result are indicated as dark and light gray (1-- and 2--$\sigma$) filled regions.}
\label{fig:limit}
\end{center}
\end{figure}

WIMP signal hypotheses are tested with a PLR statistic as in\,\cite{Akerib:2015:run3}, scanning over spin-independent WIMP-nucleon cross sections at each value of WIMP mass. Nuclear-recoil energy spectra for the WIMP signal are derived from a standard Maxwellian velocity distribution with $v_0=220$\,km/s, $v_{esc}$ = 544\,km/s, $\rho_0$ = 0.3\,GeV/cm$^3$, average Earth velocity of 245\,km/s, and a Helm form factor. Detector response nuisance parameters, describing all non-negligible systematic uncertainties in the signal and background models, are listed with their constraints and observed fit values in Table~\ref{tab:fit_parameters}. 
Systematic variation of the electric field models in the 16 exposure segments, constrained within the uncertainties of the $^3$H-based NEST model fits, results in negligible ($<$4\%) change in projected sensitivity. The likelihood is the product of terms for the full (signal plus background) PDF evaluated at each event, a Poisson term for the observed number of events, and the set of Gaussian constraints. The field-dependence of the detector response is included by treating the date bins as separate exposures, with detector response variation in drift time included in the date-bin-specific $\mathrm{\{}$\emph{S1}, \emph{S2}, $\mathrm{r_{S2}}$, $\mathrm{\phi_{S2}}$, $\mathrm{z_{S2}}\mathrm{\}}$ PDFs.

The data are in good agreement with the background-only model, having a PLR $p$ value of 0.39 at 100\,\GeVmass. Goodness of fit is also assessed by Kolmogorov-Smirnov tests for PDF projections in the five observables, which each return $p>0.6$. We present the 90\% C.L.~upper limit on cross section versus mass in Fig.~\ref{fig:limit}, as the gray curve labeled ``LUX WS2014--16''. It has a minimum of \SI{2.2e-46}{\centi\metre\squared} at \SI{50}\,\GeVmass, corresponding to 4.2 expected signal events. Compared to WS2013\,\cite{Akerib:2015:run3}, new WIMP parameter space is excluded at all masses above 7~\GeVmass, with a fourfold improvement in sensitivity for all masses above 80\,\GeVmass.

In addition to the exclusion limit from WS2014--16 data alone, we also perform an analysis which combines the WS2014--16 data with those of WS2013. This combined analysis is done by joining the event-level data sets themselves, and not by combining exclusion curves.  This is an important point, because the published WS2013 exclusion curve in\,\cite{Akerib:2015:run3} (also shown in Figure\,\ref{fig:limit}) is power constrained, due to a significant downward fluctuation in the background in that data set. Therefore the combined sensitivity is better than what might naively be expected by considering the published exclusion curves alone. The data sets are combined by treating WS2013 as a 17th exposure segment.  Since each exposure segment is given its own response, signal, and background models, this method simplifies the combination of the two data sets which have important differences.  First, WS2013 data and models use two spatial coordinates while WS2014--16 uses three.  Second, the spatial coordinates of WS2013 are corrected for nonvertical electron drifts, which is not done in WS2014--16 models and data.  Third, the WS2013 background model includes a component from $^{127}$Xe, which had decayed away by the start of WS2014--16.  Response, signal, and background models for this WS2013 exposure segment are carried over unchanged from\,\cite{Akerib:2015:run3}. Nuisance parameters described in Table\,\ref{tab:fit_parameters} are treated as independent between WS2013 and WS2014--16, with the exception of the Lindhard $k$ parameter.  We conservatively apply a power constraint\,\cite{Cowan:2011an} at the $-1\sigma$ extent of the projected sensitivity in order to avoid excluding cross sections for which the sensitivity is unreasonably enhanced through chance background fluctuation. The combined 90\% C.L.~upper limit is shown as the thick black curve in Fig.\,\ref{fig:limit} labeled `LUX WS2013+WS2014--16'.  This combined exclusion limit reaches a minimum of \SI{1.1e-46}{\centi\metre\squared} at 50\,\GeVmass, corresponding to an expected 3.2 signal events. This significant advance has newly tested some of the most favored WIMP parameter space, including models consistent with the SUSY CMSSM as plotted in Fig.~\ref{fig:limit}.

This work was partially supported by the U.S. Department of Energy (DOE) under Award No.~DE-AC02-05CH11231, No.~DE-AC05-06OR23100, No.~DE-AC52-07NA27344, No.~DE-FG01-91ER40618, No.~DE-FG02-08ER41549, No.~DE-FG02-11ER41738, No.~DE-FG02-91ER40674, No.~DE-FG02-91ER40688, No.~DE-FG02-95ER40917, No.~DE-NA0000979, No.~DE-SC0006605, No.~DE-SC0010010, and No.~DE-SC0015535; the U.S. National Science Foundation under Grants No.~PHY-0750671, No.~PHY-0801536, No.~PHY-1003660, No.~PHY-1004661, No.~PHY-1102470, No.~PHY-1312561, No.~PHY-1347449, No.~PHY-1505868, and No.~PHY-1636738; the Research Corporation Grant No.~RA0350; the Center for Ultra-low Background Experiments in the Dakotas (CUBED); and the South Dakota School of Mines and Technology (SDSMT). LIP-Coimbra acknowledges funding from Funda\c{c}\~{a}o para a Ci\^{e}ncia e a Tecnologia (FCT) through the Project-Grant No.~PTDC/FIS-NUC/1525/2014. Imperial College and Brown University thank the UK Royal Society for travel funds under the International Exchange Scheme (No.~IE120804). The UK groups acknowledge institutional support from Imperial College London, University College London and Edinburgh University, and from the Science \& Technology Facilities Council for PhD studentships No.~ST/K502042/1 (AB), No.~ST/K502406/1 (SS) and No.~ST/M503538/1 (KY). The University of Edinburgh is a charitable body, registered in Scotland, with Registration No.~SC005336.

This research was conducted using computational resources and services at the Center for Computation and Visualization, Brown University, and also the Yale Science Research Software Core. The $^{83}$Rb used in this research to produce $^{83\mathrm{m}}$Kr was supplied by the United States Department of Energy Office of Science by the Isotope Program in the Office of Nuclear Physics.

We gratefully acknowledge the logistical and technical support and the access to laboratory infrastructure provided to us by SURF and its personnel at Lead, South Dakota. SURF was developed by the South Dakota Science and Technology Authority, with an important philanthropic donation from T. Denny Sanford, and is operated by Lawrence Berkeley National Laboratory for the Department of Energy, Office of High Energy Physics.

\bibliography{main}

\clearpage
\begin{widetext}
\appendix
\section{Supplementary Material}
\label{sec:appendix_section}

The simulated 3-D electric-field maps in WS2014--16 are constructed by comparing the spatial distribution, in observed coordinates, of a physically uniform calibration source ($^{83\mathrm{m}}$Kr) to that predicted by the field model.  The observed coordinates measure the electron drift time and the $x$-$y$ position of electrons as they leave the liquid surface.  Figure\,\ref{fig:KrWallRadius} shows one method of comparing model to data (see caption).

\begin{figure}[ht!]
\begin{center}
\includegraphics[width=0.75\textwidth,clip]{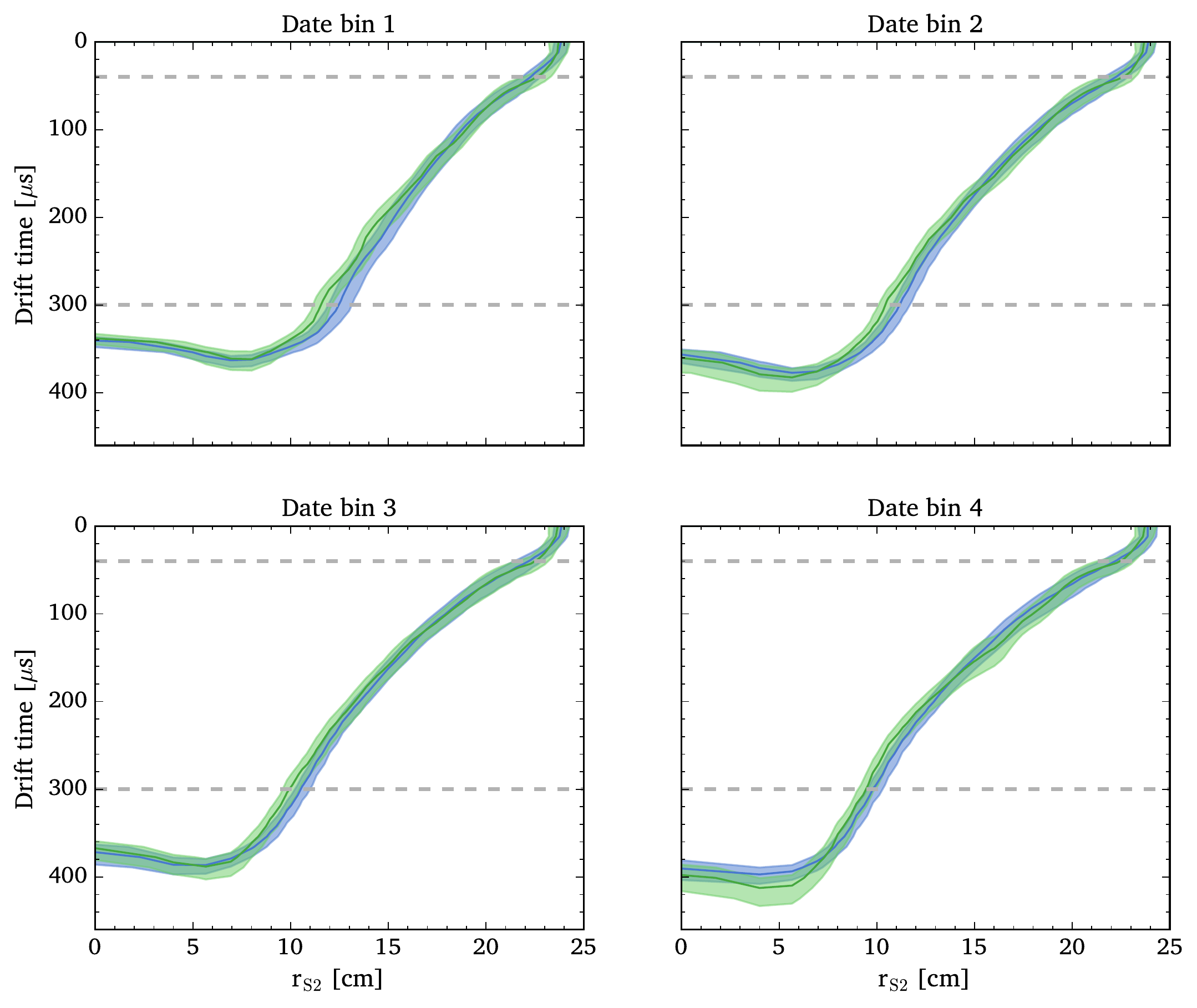}
\caption{A comparison of the measured position of the detector wall and cathode to that predicted by the best-fit electrostatic field model.  As the electrons are drifted upwards, they are pushed radially inwards; they therefore exit the liquid surface (where they are detected and their $x$-$y$ position is measured) at a radius that is less than the radius at which they originated.  As a result, the measured shape of the detector wall, which is physically vertical, is warped in observed coordinates.  Similarly, though the cathode is physically horizontal, the field-dependent drift velocity of electrons in liquid xenon causes its shape to appear as an inverted `U' in measured coordinates.  In each of the four axes, the blue contour is the measured shape of the detector wall from calibration data, while the green contour indicates the prediction of the wall shape from the best-fit field model.  The width of each contour indicates the uncertainty in the wall position resulting from the histogram bin sizes used to construct the contours. Note that the radius of the wall in observed coordinates (``r$_{\mathrm{S2}}$'') is not axially symmetric, and therefore the contours here represent an average over azimuthal angle (this is not the fit space; the fits are instead performed in 3-D).  The background model for events from radon plate-out on the walls is constructed directly in measured coordinates entirely from side bands, and does not use these field maps. Horizontal gray-dashed lines, at 40 and 300\,$\mu$s, indicate the drift-time extent of the fiducial volume used in WS2014--16.}
\label{fig:KrWallRadius}
\end{center} 
\end{figure}

\end{widetext}

\end{document}